\documentclass[aps,pre,twocolumn,amssymb,showpacs,groupedaddress]{revtex4-2}
\usepackage{graphicx,color}
\usepackage[normalem]{ulem}
\usepackage{amsmath}
\usepackage{tabularx}
\def\beq{\begin{equation}}
\def\eeq{\end{equation}}
\def\bea{\begin{eqnarray}}
\def\eea{\end{eqnarray}}

\definecolor{colortodo}{RGB}{255,0,0}

\definecolor{colortodo2}{RGB}{0,150,0}

\begin{document}
\title{Dynamics of magnetoelastic robots in water-saturated granular beds}
\author{Animesh Biswas, Trinh Huynh, Balaram Desai, Max Moss, and Arshad Kudrolli}
\affiliation{Department of Physics, Clark University, Worcester, Massachusetts 01610, USA}
\date{\today}

\begin{abstract}
We investigate the dynamics of a magnetoelastic robot with a dipolar magnetic head and a slender elastic body as it performs undulatory strokes and burrows through water-saturated granular beds. The robot is actuated by an oscillating magnetic field and moves forward when the stroke amplitude increases above a critical threshold. By visualizing the medium, we show that the undulating body fluidizes the bed, resulting in the appearance of a dynamic burrow, which rapidly closes in behind the moving robot as the medium loses energy. We investigate the applicability of Lighthill's elongated body theory of fish locomotion, and estimate the contribution of thrust generated by the undulating body and the drag incorporating the granular volume fraction-dependent effective viscosity of the medium. The projected speeds are found to be consistent with the measured speeds over a range of frequencies and amplitudes above the onset of  forward motion. However, systematic deviations are found to grow with increasing driving, pointing to a need for further sophisticated modelling of the medium-structure interactions. 
\end{abstract}

\maketitle
\section{Introduction}
The principles of locomotion in water-saturated loose sediments are important to understanding organisms that inhabit the bottom of ponds, lakes, and oceans \cite{barnes2009fundamentals}. A great variety of fish, crustaceans, echinoderms, among other organisms live and burrow into the sandy beds under water to prey, or escape currents. Among the various body strokes observed in organisms moving in Newtonian and non-Newtonian fluids~\cite{lauga09,maladen09,hosoi15,maladen2011mechanical,shimada2009swimming,ruhs2021complex}, undulatory strokes have been found to be effective in loosely consolidated granular medium at shallow depths~\cite{gray53,jung10,dorgan13,dorgan15,Kudrolli2019,herrel2011burrowing,yaqoob2023mechanics}. Indeed, undulatory motion is used by sperm, eels, and nematodes to move through Newtonian and non-Newtonian mediums such as mucus and mud~\cite{Fischer2022, suarez2003hyperactivated, spagnolie2023swimming}. These observations have been used to design robots which mimic fish swimming in water~\cite{Ramananarivo2013,wang2021effect} to snakes and earthworms burrowing in loose sand~\cite{Naclerio2021,Das2023,naclerio2018soft}. 

In the loose granular beds of interest, the grains themselves are athermal, and rapidly come to rest unless actively agitated~\cite{mehta2007granular}. Thus, granular sediments immersed in a fluid show a yield-stress and shear-thinning behavior~\cite{balmforth14,panaitescu17}, which is quite different from the limits of Newtonian fluids and dry granular beds where most studies on locomotion have been carried out. The fluidization of the medium by the moving intruder modifies the packing and rheology of the medium around the body~\cite{nichol2010flow}. While moving with a constant velocity, the drag of an intruder scales linearly with overburden pressure due to the weight of the granular medium above~\cite{panaitescu17,jewel18}. However, the drag does not increase as rapidly when the intruder speed increases and thus fluidization increases~\cite{allen19}, or when wake interactions become more important~\cite{Chang2022}. These issues are compounded in bodies that use rapid undulatory strokes in sediments immersed in a fluid. Because the motion of the body and the medium are intimately intertwined, having a complementary view of both the structure and the medium is important in gaining a full understanding of the dynamics.

Here, we construct biomimetic magnetoelastic robots to investigate locomotion in granular sediments immersed in water. We perform experiments to study the form and dynamics of an undulating robot with a magnet head and elastic body in a fluid-saturated granular media driven by an oscillating magnetic field. In contrast with previous studies of locomotion in granular beds which are in the quasi-static frictional regime, we focus here on the inertial hydrodynamic regime where the medium is highly fluidized as a consequence of the burrowing. Exploiting refractive index matching of the grains, we measure the shape and speed of the robot while it moves through the medium. Further, we visualize the resulting flow of the medium using fluorescence techniques to gain a deeper understanding of the hydrodynamic interactions experienced by the moving robot. We find that the body strokes strongly fluidize the medium resulting in a suspension with a granular volume fraction dependent viscosity. We then discuss the observed robot locomotion speeds in terms of Lighthill's elongated body theory~\cite{lighthill1970} of fish locomotion.  

\section{Experimental methods}
 \begin{figure*}
 \begin{center}
  	\includegraphics[width=14 cm]{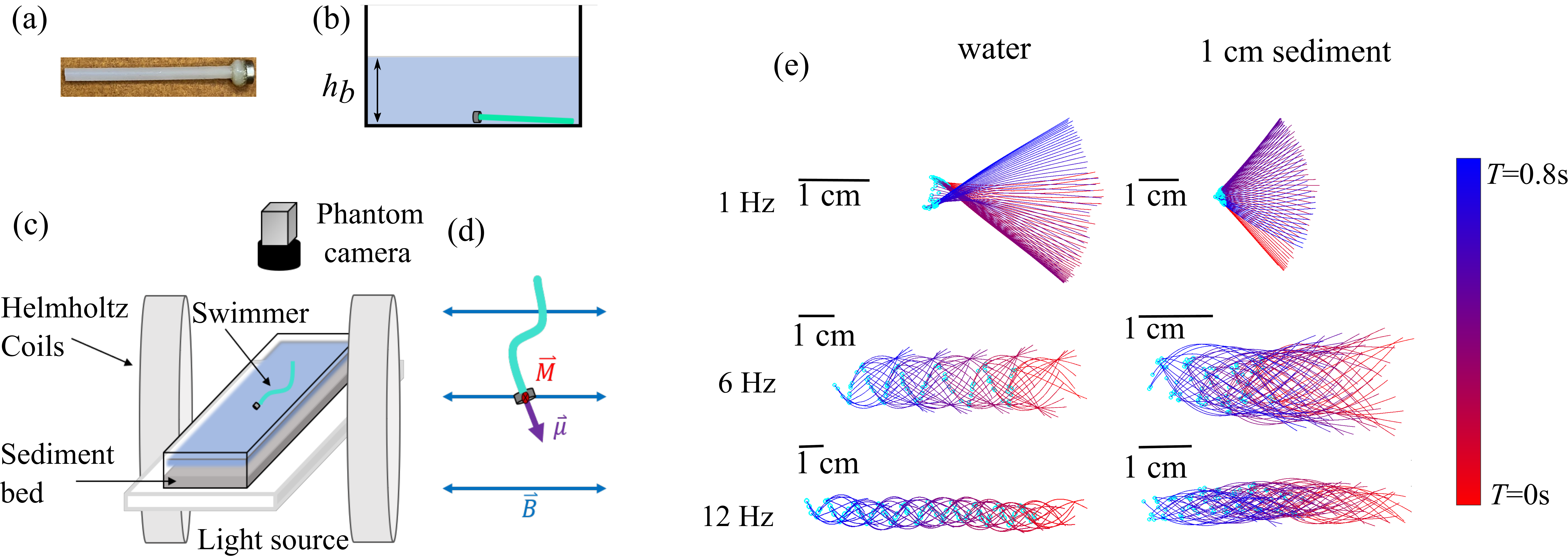}\\
	\end{center}
	\caption{(a) Image of the magnetoelastic robot with a magnet head and elastic body. (b) Schematic of the side view of the system showing the robot placed at the bottom of the container with sediment bed. $h_b$ is the sediment bed height. (c) Schematic of the experimental system shows Helmholtz coils, light source, camera, robot in the container. (d) The magnetic field $\vec{B}(t)$ results in a torque $\vec{M}(t) = \vec{\mu} \times \vec{B}(t)$ which oscillates periodically. (e)  The tracked body shapes of the robot moving in water ($h_b= 0$\,cm), and in a sediment bed ($h_b = 1$\,cm) at various frequencies. These representative tracked body shapes clearly illustrate that the robot travels greater distances when swimming in water as opposed to the granular hydrogel medium in the case of 6\,Hz and 12\,Hz. Furthermore, these tracks demonstrate the relative stability of the robot's direction of motion. The snapshots are plotted in 10\,ms time intervals over the time denoted in the color bar.}
	\label{fig:setup}
\end{figure*} 

An image of the magnetoelastic robot used in our experiments is shown in Fig.~\ref{fig:setup}(a), a schematic of the experimental cell with robot can be seen in Fig.~\ref{fig:setup}(b) where the robot is placed at the bottom of the cell, and a schematic of the experimental system can be found in Fig.~\ref{fig:setup}(c). The robot is constructed by attaching a small cylindrical Neodymium magnet of mass $m_m= 77$\,mg, diameter 3\,mm, and height 2\,mm, to the end of an elastic polyvinylsilioxane rod with density $\rho_r = 980$\,kg\,m$^{-3}$, diameter $d = 1.63$\,mm, and Young's modulus $E = 0.35$\,MPa. The diameter of the elastic body near the head is increased to approximately $2d$ in order to attach the magnet firmly, resulting in a total robot mass $m=0.164$\,g, length $L=32$\,mm. The granular media used in our investigations consists of hydrogel beads with diameter $d_g$ ranging between 0.5\,mm and 2\,mm, and density $\rho_g = 1004$\,kg\,m$^{-3}$ while immersed in water with density $\rho_w = 997$\,kg\,m$^{-3}$ at $23^\circ$C. This granular medium is chosen because it is essentially refractive index matched with water enabling us to fully visualize the robot in situ.  The resulting sediment bed is known to be described by the Herschel–Bulkley stress-strain model under steady driving conditions~\cite{panaitescu17,pal2021}, with a yield stress which increases linearly with depth in the sediment.

We study the dynamics of the robot as a function of increasing bed height $h_b = 1$\,cm, 2\,cm, and 3\,cm, and contrast with swimming dynamics in water, i.e. when $h_b =0$\,cm. The container is placed at the center of a Helmholtz coil. An alternating current is passed through the coil which results in a magnetic field $B(t) = B_o \cos(2\pi f t)$, where $B_o$ is the amplitude of the sinusoidal magnetic field, $f$ is its oscillation frequency, and $t$ is time. The field is observed to be spatially uniform within 2.2\%, where measurements are made. 
The magnet has a magnetic moment ${\mu} = 2.1 \times 10^{-3} 
$\,A\,m$^2$ that results in a torque $\vec{M}(t) = \vec{\mu} \times \vec{B}$, where $\vec{B} = B(t) \hat{y}$, causes the magnetic head of the robot to rotate as illustrated in Fig.~\ref{fig:setup}(d). This applied torque leads the elastic body to rotate and bend depending on its elasticity, and the drag exerted by the medium. 

The robot is placed in a transparent acrylic container with the granular bed, and water is filled well above the bed surface to avoid any capillary effects (Fig.~\ref{fig:setup}(b)). Because the magnetic head has a higher density compared with the medium, the robot sinks to the bottom of the bed, and stays in apparent contact with the solid substrate even while it moves. The top view of the robot's motion can be seen in Movie~S1 in Supplemental Material\cite{Supplemental} and an example video in Movie~S4\cite{Supplemental} shows the robot's motion from a side view. The refractive index of the grains is approximately the same as water, enabling us to image the robot in real-time as the robot moves inside the medium with back-lighting generated with a LED array. We use a Phantom model VEO-E 310L camera with a pixel resolution of $1024 \times 512$ at 50 or 100 frames per second depending on the applied frequency and speed of the robot to image the robot dynamics (see Movie~S1 in Supplemental Material\cite{Supplemental}). We observe that the elastic body undulates in the horizontal plane, perpendicular to the torque acting on the magnetic head with the same frequency as the oscillating magnetic field. We track the robot position and shape in each frame with algorithms using the image processing toolbox in MATLAB~\cite{Kudrolli2019} to further analyze its dynamics. 

\begin{figure}
	\includegraphics[width=9 cm]{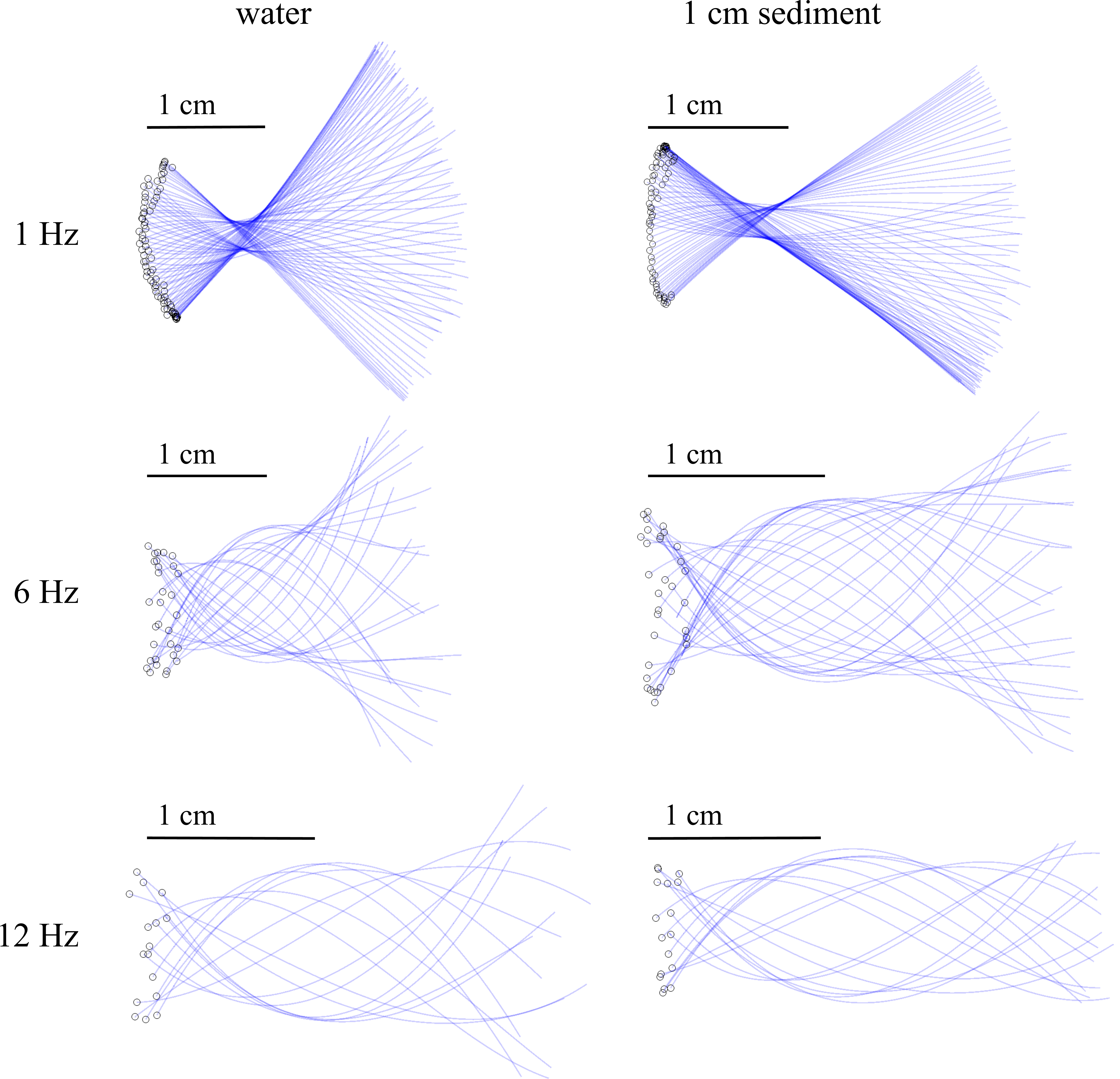}
	\caption{The superimposed snapshots of tracked body shapes of the robot moving in water ($h_b= 0$\,cm), and in a sediment bed ($h_b = 1$\,cm) at various frequencies. These snapshots are plotted at 10 ms time intervals over a single cycle of body oscillation in the co-moving reference of the robot center of mass. These snapshots portray the evolving body shapes during motion and demonstrate the transition from rigid body motion to anguilliform motion as a function of frequency.}
	\label{fig:snapshot}
\end{figure} 

\section{Observed Strokes and Speeds} 

\begin{figure}
	\includegraphics[width=9 cm]{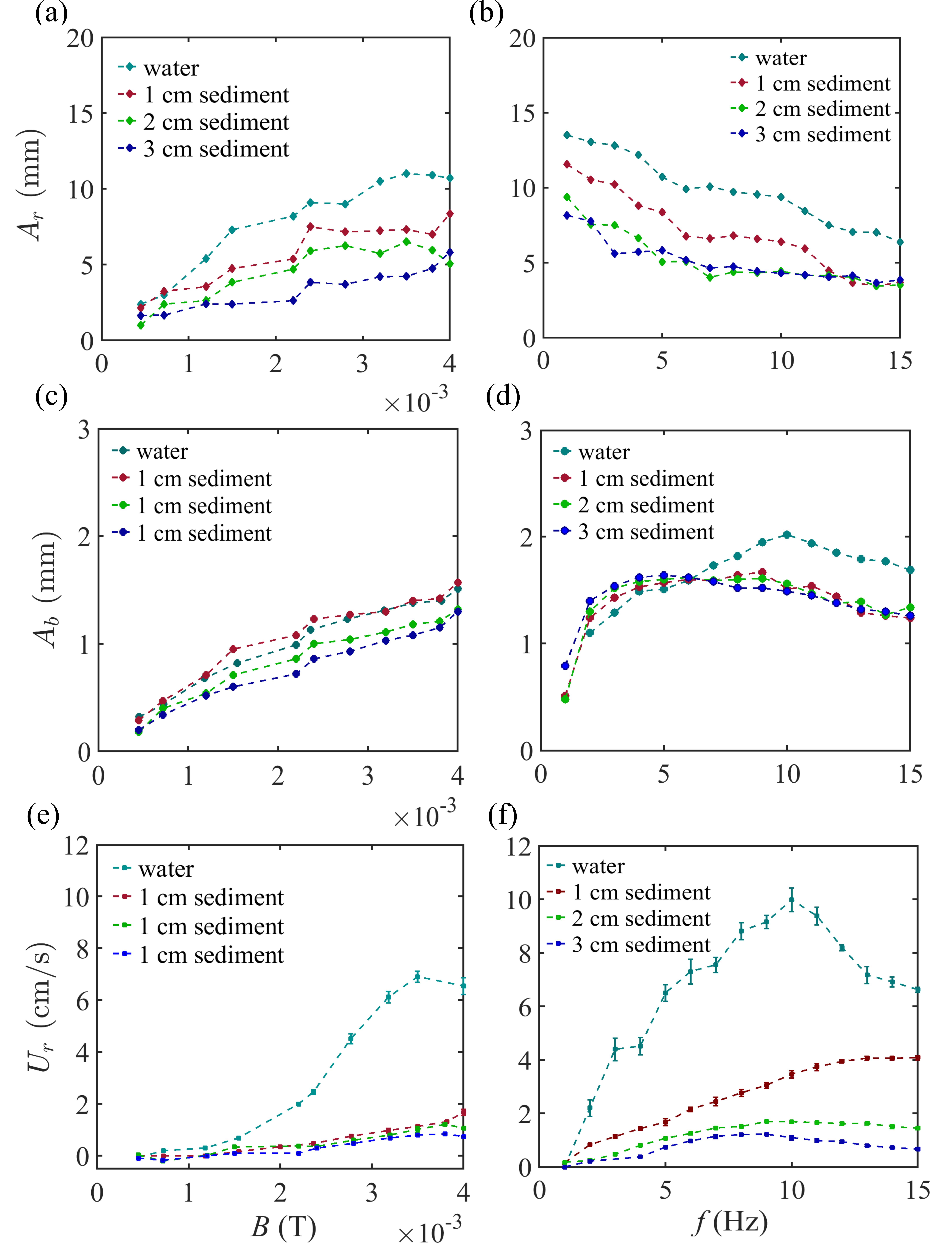}
	\caption{The tail oscillation amplitude $A_{r}$ of the magnetoelastic robot in water and in sediments as a function of field strength (a), and frequency (b).  The corresponding deformation amplitude $A_b$ (c), (d), and speed of the robot $U_r$ (e), (f). The frequency is $f = 5$\, Hz when the field $B$ is varied, and the field is $B_o = 4$\,mT when the frequency $f$ is varied.}
	\label{fig:speed}
\end{figure}

Figure~\ref{fig:setup}(e) shows representative forms of the robot as it moves in water and in sediment bed at various driving frequencies. While the body oscillates with large amplitudes, the robot is not observed to move forward at the lowest frequency. At higher frequencies, the robot is observed to move forward with increasing speed, while the body stroke amplitude decreases. The stroke shapes can be observed to be periodic, and the direction of motion is quite stable even though it is not actively controlled. The plotted overlapping body shapes illustrates that the robot moves forward  qualitatively faster in water than in the granular medium. Figure~\ref{fig:snapshot} illustrates the superimposed body strokes observed in the co-moving reference of the robot center of mass. The snapshots are plotted both in water and in sediment at 10 ms time intervals over a single period of body oscillation at the three different frequencies. The distinct shapes of the body strokes can be seen in this figure. Comparing the robot forms and displacements in the water and sediments, we note that they are systematically lower in the sediments, indicating that the medium exerts higher drag under otherwise similar applied driving conditions. 

We plot the transverse oscillation amplitude of the body at the tail $A_r$ as a function of applied magnetic field and frequency in Fig.~\ref{fig:speed}(a) and (b), respectively, and the bending amplitude of the body $A_b$ corresponding to the mean square deviation from a straight line fit to the body in Fig.~\ref{fig:speed}(c) and (d). $A_b$ demonstrates how body strokes transition from rigid body rotation to an anguilliform pattern as a function of $f$, and $B$, as also depicted in Fig.~\ref{fig:snapshot}. As may be expected, we observe that $A_r$ and $A_b$ both increase with $B$, as greater torque is applied to the head by increasing the magnetic field. On the other hand, $A_r$ is observed to decrease with increasing $f$, while $A_b$ increases, reaches a peak, and then decreases with increasing $f$. These trends are consistent with the fact that drag acting on a body increases with its speed. Nonetheless, the change in strokes may be characterized as evolutionary as the robot burrows in the sediment bed versus swimming in water.

We obtain the mean speed of the robot $U_r$ by measuring the time required to travel a distance of at least 5\,cm, or by the measuring the distance moved over 5 to 10\,seconds in cases where the robot moves very little. The error bars correspond to mean square deviations from means obtained over 5 trials. Fig.~\ref{fig:speed}(e, f) show plots of $U_r$ as a function of magnetic field $B$ and frequency $f$, respectively. As $B$ is increased, we observe that the robot moves forward with increasing speeds above $B_o \approx 0.5$\,mT, corresponding to a sufficiently large $A_r$. The speeds in the sediments are systematically lower compared with the speeds in water. Similarly with increasing oscillation frequency, we observe that the robot moves forward only above a critical frequency $f > 1$\,Hz with increasing speed. Again, the speeds can be observed to be lower in the sediments and decrease systematically with the increase in $h_b$. However, $U_r$ can be observed to increase and then decrease with frequency, with the peak occurring at lower frequencies with increasing $h_b$. Thus, a complex response can occur depending on the applied driving conditions, and interaction of the robot body with the medium.

\section{Medium fluidization}
\begin{figure}
	\includegraphics[width=8cm]{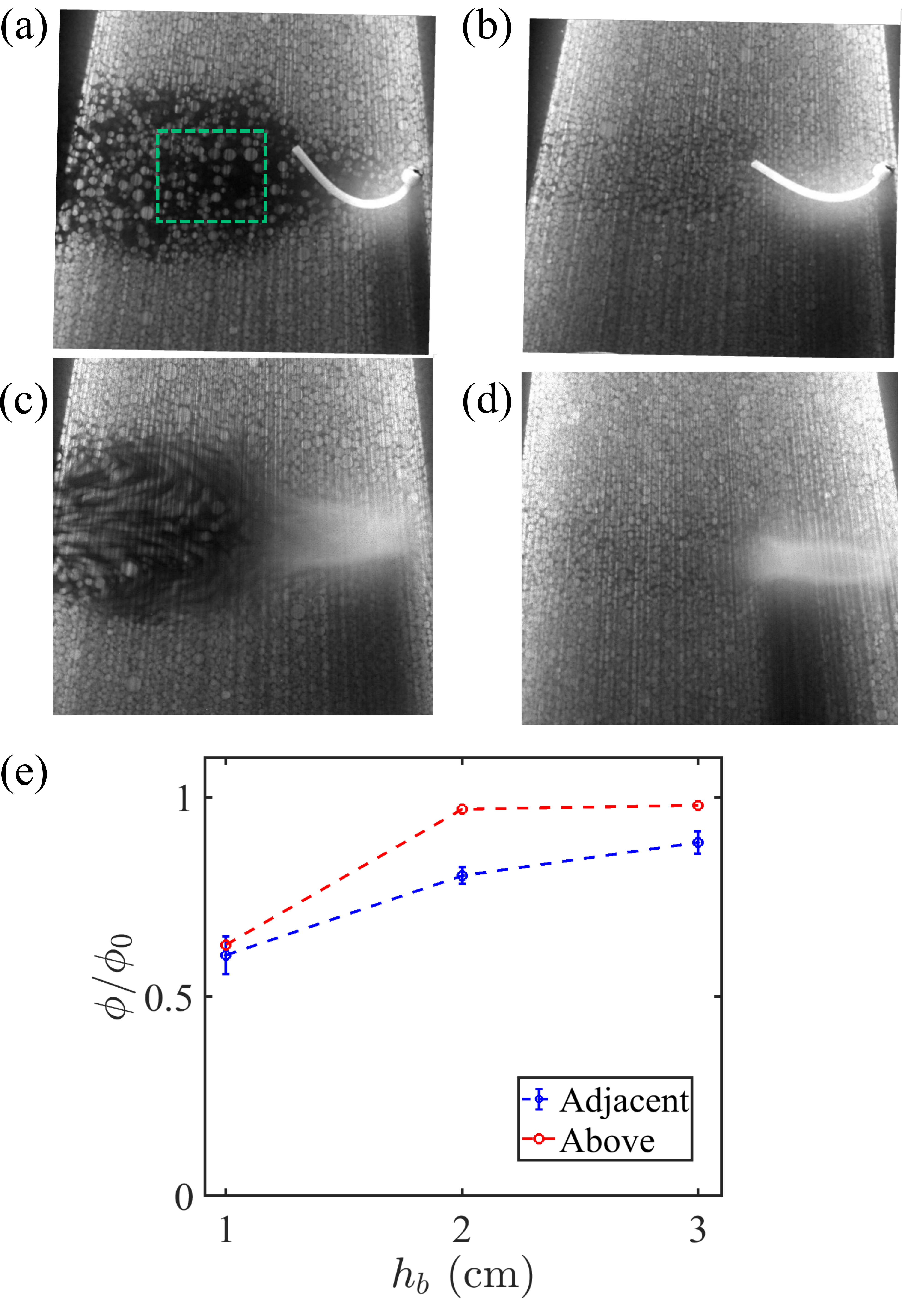}
	\caption{(a, b) Horizontal cross sections of the sediment medium in the plane of the undulating robot  visualized by illuminating the fluorescent grains with a laser sheet for (a) $h_b= 1$\,cm  and (b) $h_b=3$\,cm ($f=6$\,Hz, $B_o$=4 mT). (c, d) Images corresponding to an exposure over 3\,periods shows the flow around the robot. (c) A large region behind the robot is observed to be highly fluidized ($h_b= 1$\,cm). (d) The fluidized region is more localized around the robot moving in deeper bed ($h_b=3$\,cm). (e) The volume fraction $\phi$ is measured in the region denoted by the dashed line box (shown in (a)) at adjacent plane to the robot, and in a plane 5\,mm above the plane of the robot. The volume fraction is normalized by the volume fraction of the unfluidized sedimented bed $\phi_o = 0.6$.}
 \label{fig:medium} 
\end{figure}

Towards understanding the effect of the moving robot on the medium, we visualize the medium in the horizontal plane in which the robot moves. For these studies, the hydrogel beads are prepared by adding a small amount of dye (Rhodamine 6G) to the water in which they are hydrated. The dye stays in the beads after being placed in the clear water. Thus, the beads fluoresce when illuminated with a 532\,nm green laser sheet, and appear bright in contrast with the water in the interstitial space. Figure~\ref{fig:medium}(a) and (b) show snapshots while the robot is burrowing in sediments with $h_b=1$\,cm and $h_b=3$\,cm, respectively. In order to illustrate the fluidization and flow of the medium, time-averaged images are also shown in Figure~\ref{fig:medium}(c) and (d). The corresponding movies can be found in the Supplemental Material (S2, S3) \cite{Supplemental}. The medium can be noted to be highly fluidized near the body, and the robot appears to move through a dilute suspension in the central portion of its oscillation cycle in the case of the shallower bed. A burrow forms as the robot sweeps the granular material side to side to an extent depending on $h_b$. Clearly, the greater the overburden pressure in deeper beds, the medium fills back in more rapidly, leading to the burrow to be less pronounced.     

Figure~\ref{fig:medium}(e) shows the average volume fraction of the granular medium $\phi$ in the adjacent plane of the moving robot and in the plane above 5 mm from the robot's position. These values are obtained by counting the number of grains in a given cross sectional area (dashed line box in Figure~\ref{fig:medium}(a)), and comparing the value obtained for a well settled granular bed $\phi_o$, which is known as correspond to $\phi = 0.6$ in the case of these nearly spherical granular hydrogels~\cite{panaitescu17}. We observe that $\phi$ is well below $\phi_o$ behind the robot especially in case of shallow sediment depths. 

Thus, the picture which emerges from the visualization of the granular medium is that significant variations of packing fraction occur due to fludization of the grains because of the undulatory strokes performed by the robot. The resulting lowering of the volume fraction of the granular phase means that the effective medium encountered by the oscillating body is different along its length, and varies within the oscillation phase. In the following, we treat these differences by considering the drag encountered by the head separately from the rest of the body. 

\section{Force Analysis}

The robot moves when the thrust $T_r$ generated by the undulating filament body exceeds the drag exerted by the medium and the substrate, i.e.
\begin{equation}
T_r \ge D_h + D_b + D_s,
\label{eq:fbalance}
\end{equation}
where $D_h$ is the drag acting on the head due to the medium, $D_b$ is the drag on the body, and $D_s$, the drag due to the friction from substrate. To evaluate these terms, we have to examine the hydrodynamic regime.

\begin{figure}
    \centering
 
    \includegraphics[width=8.5cm]{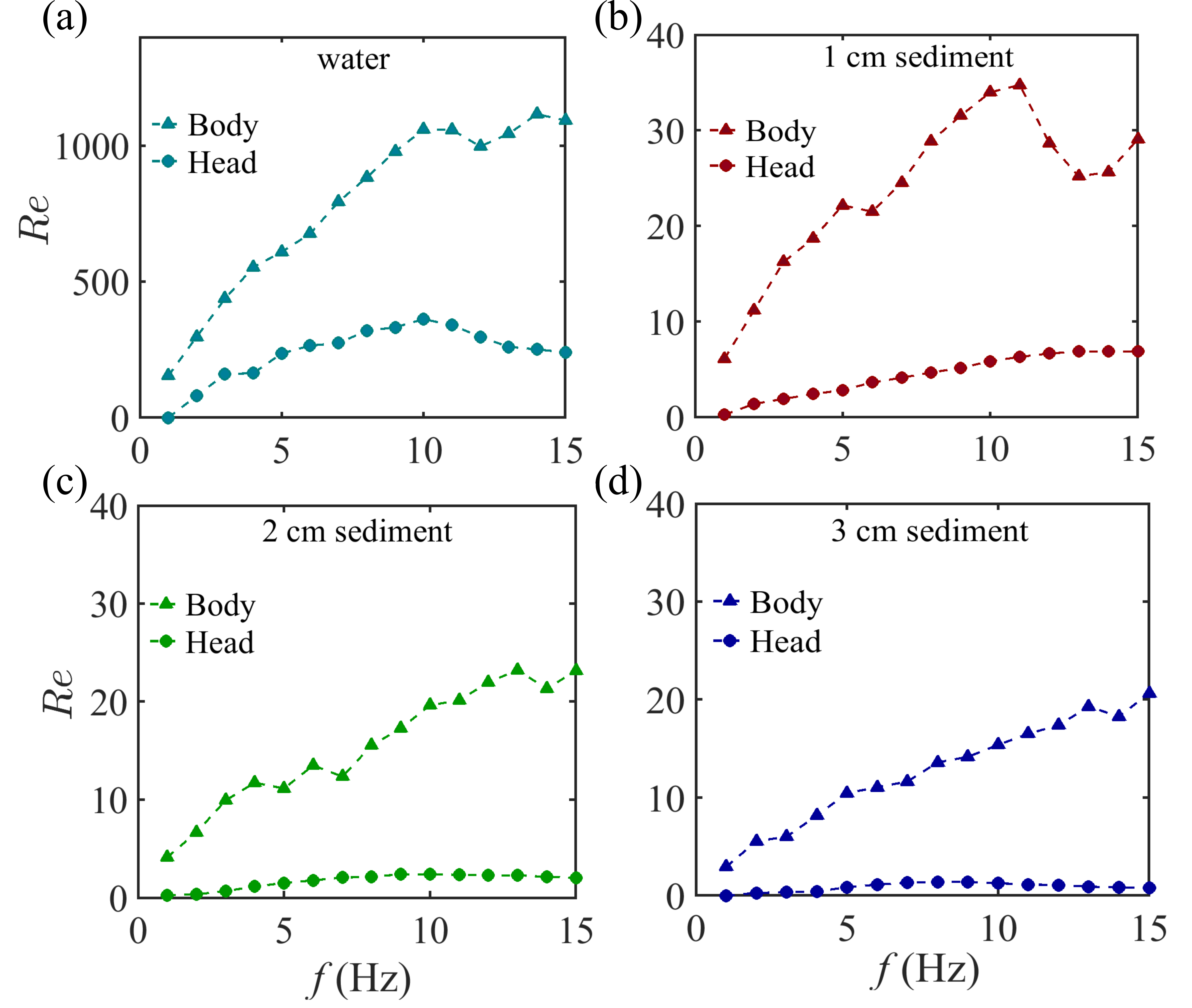}
    \caption{The Reynolds number corresponding to the robot's body and head in water (a) and sediments with $h_b = 1$\,cm (b), $h_b = 2$\,cm  (c), $h_b = 3$\,cm  (d).}
    \label{fig:Re}
\end{figure}

We evaluate the Reynolds number $Re = {\rho U_s L_s}/{\eta_m}$, where $\rho$ is the density of the medium, $\eta_m$ is the viscosity of the medium, $U_s$ the velocity over the length scale $L_s$ to determine the relevant hydrodynamic regime. In the case of water, $\rho \approx 1000$\,kg/m$^3$, and $\eta_w \approx 1.0$\,mPa.s at $23^\circ$C. Then, over the scale of the robot's head, $L_s = 2d$, and $U_s = U_r$, and for considering the lateral motion of the body $U_s = 2\pi A_{r} f$. Figure~\ref{fig:Re}(a) shows a plot of $Re$ over the range of $f$ in the case of water. While $Re$ is much greater in the case of the body compared with the head, and in both cases $Re \gg 1$,  Thus, the robot is in the inertia dominated regime in water. 

In evaluating $Re$ in the case of the sediments, it is clear from Fig.~\ref{fig:medium}(a, b) that the body encounters a  granular medium which is well fluidized. Because of the nonuniform spatial and temporal nature of the suspension, 
it becomes difficult to evaluate $\eta_m$. Nonetheless, given that $A_{r}$ and $U_r$ are only about a factor of 2 lower in the sediments, one may anticipate that the hydrodynamic regime is important to propulsion may also be inertial in the case of the sediments.  We proceed with this assumption to evaluate the components need to achieve locomotion, and check for self consistency at the end.   

\begin{figure*}
    \centering
    \includegraphics[width=14cm]{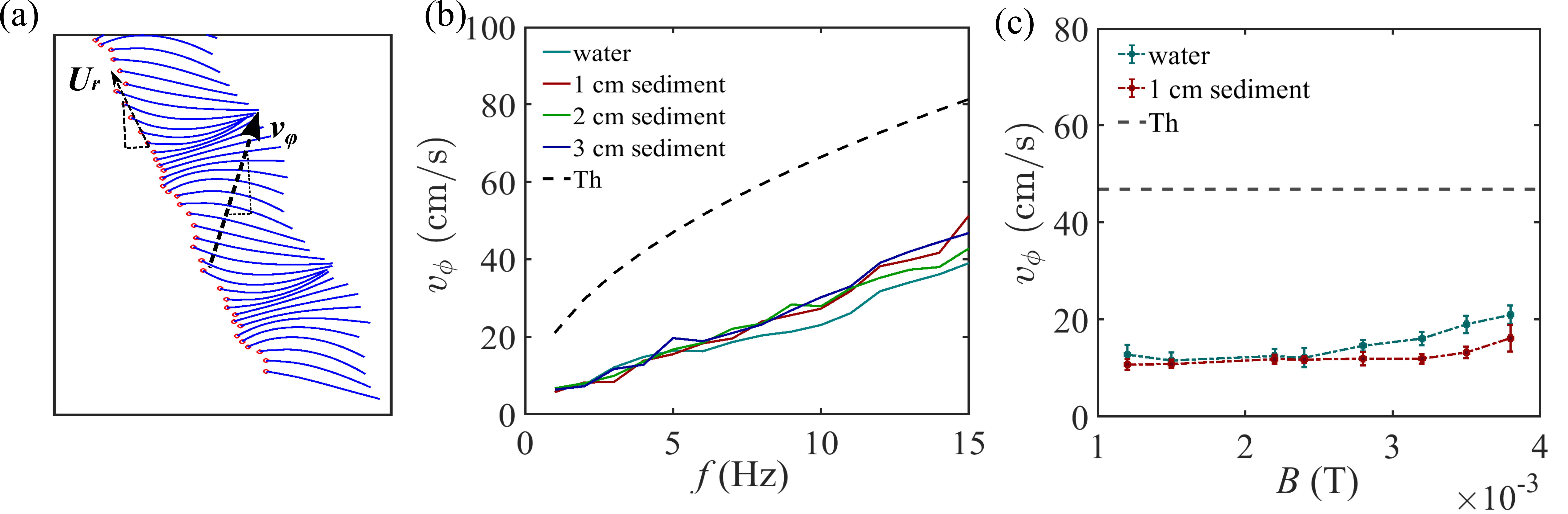}
    \caption{(a) Curtain plot of robot shapes to illustrate the measurement of the phase velocity $v_\phi$ and swimming velocity $U_r$. The snapshots are plotted in 10 ms time intervals. The measured $v_\phi$ are observed to be overall similar within the experimental measurement errors for varying frequency $f$ (b) and varying $B$ (c). The measured values of $v_\phi$ are compared with the Euler-Bernoulli beam theory (dashed line, Eq.~(\ref{eq:vphi})).}
    \label{fig:vphi}
\end{figure*}

According to Lighthill's theory of anguilliform swimming in the inertial regime, the thrust calculated at the tail position is given by~\cite{lighthill1970,Ramananarivo2013},
\begin{equation}
     T_r = \frac{1}{2} \rho S [ \langle (\partial_t y)^2\rangle] - U_r^2 \langle (\partial_x y)^2\rangle]_{tail},
\end{equation}
where $\rho$ is the fluid density, and $S = \pi d^2/4$ the body cross section.
If the body undulations are described by
\begin{equation}
    y(x,t) = A_r \cos(2 \pi (ft - x/\lambda)),
    \label{eq:y_cos}
\end{equation}

where $A_r$ is the stroke amplitude, and $\lambda$ is the wavelength. After substituting, and since $v_\phi = \lambda f$,
we have 
\begin{equation}
    T_r = {\pi^2 f^2 \rho S A_r^2} [ 1 - (\frac{U_r}{v_\phi})^2 ].
    \label{eq:thrust_alt}
\end{equation}
Besides the oscillation frequency, amplitude, and swimming speed, we note that the phase velocity of the traveling wave along the body plays an important role in determining the swimming speed~\cite{lighthill1970,lighthill76}. 

\vspace{0.1in}

\begin{figure}
    \centering
   \includegraphics[width=8.5cm]{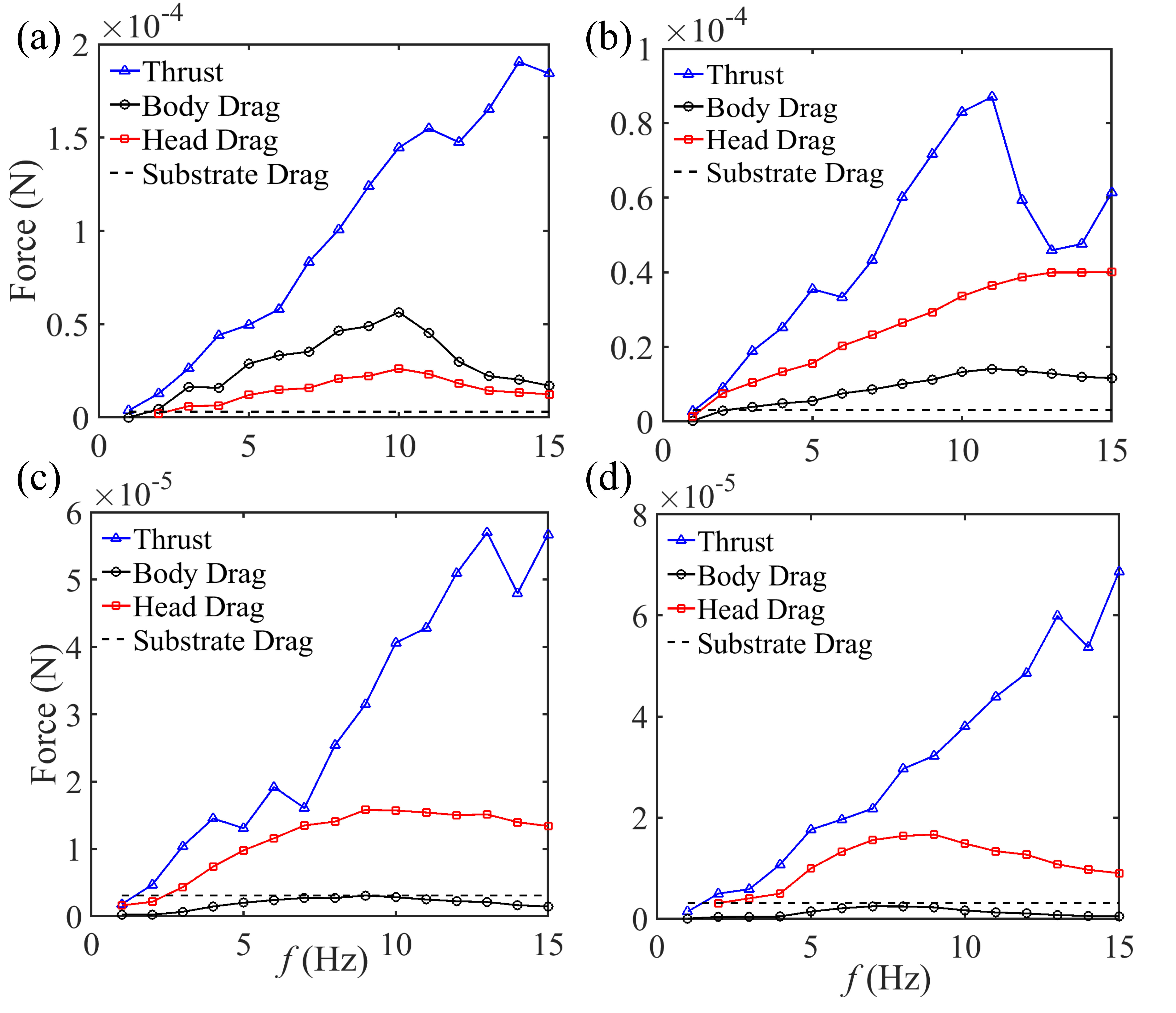}
    \caption{The estimated thrust and drag components for robot moving in water (a) and in sediments with various $h_b$= 1 cm sediment (b), 2 cm sediment (c), and 3 cm sediment (d).}
    \label{fig:Licomp}
\end{figure}

Figure~\ref{fig:vphi}(a) illustrates the traveling wave, where we determine the velocity of the traveling wave $v_\phi$ by measuring the distance between the antinodes over a given time interval. While the shape is not described by a purely sinusoidal form, we have found that this method gives a reasonable description of the overall shape while estimating using $v_\phi$/$f$. The plot of measured phase velocity $v_\phi$ versus frequency $f$ is shown in Fig.~\ref{fig:vphi}(b), and plot of measured $v_\phi$ versus field $B$ is shown in Fig.~\ref{fig:vphi}(c). According to the Euler-Bernoulli beam theory~\cite{Wang2016}, the phase velocity of a small amplitude sinusoidal wave traveling in an undamped beam, is given by 

\begin{equation}
    v_\phi = \Big(\frac{E I}{ \rho A} \Big)^{1/4} \sqrt{2\pi f}
    \label{eq:vphi}
\end{equation}
where, $E$ is Young's modulus, $\rho$ is the density of the robot material, $A$ is the cross section of the robot body, $f$ is the frequency of the oscillation, $I$ is moment of inertia. In comparing with the data shown in Fig.~\ref{fig:vphi}(b), we observe that while the estimated speed is of the right order of magnitude, it is systematically higher by a factor of about 2 than the measured speed. Further, we observe from Fig.~\ref{fig:vphi}(c) that measured $v_\phi$ increases with applied $B$, which results in higher $A_r$. We ascribe these differences to the fact that the undulatory form of the robot cannot be described by a simple sinusoidal form, and the medium also induces damping. Thus, we use the measured $v_\phi$, rather than Eq.~(\ref{eq:vphi}) to evaluate thrust. Figure~\ref{fig:Licomp}(a) shows a plot $T_r$ as a function of $f$ using the measured values of $A_r$, $U_r$, and $v_\phi$ in water. Thrust can be noted to increase somewhat linearly with increasing $f$, even as $A_r$ decreases. 

Now, the drag due to the head, 
\begin{equation}
D_h = \frac{1}{2} C_{D} \rho_w S_h U_r^2,    
\end{equation}
where $C_D$ is the drag coefficient of the head, and $S_h$ is the cross sectional area of the head. 
And, the drag due to the body alone is,
\begin{equation}
    D_b = \frac{1}{2} \rho C_D  S' U_r^2,
\end{equation}
where $C_D$ is the drag coefficient of the body, and $S' = 2 A_r d$~\cite{Ramananarivo2013}. Given $Re$ is of $O(10^2)$ in our system, we assume a simplified form $C_D \approx 24/Re + 0.35$~\cite{tritton59} to minimize the number of fitting parameters, and plot $D_h$ and $D_b$ in the case of water in Fig.~\ref{fig:Licomp}(a).

Now, considering the frictional interaction between the robot and the substrate, and because the density of polyvinylsilioxane is essentially the same as water and the sediments we have, 
\begin{equation}
    D_s = \mu_k W_b,
    \label{eq:Ds}
\end{equation}
where $\mu_k$ is the coefficient of kinetic friction of the polyvinylsilioxane head sliding on the acrylic substrate, $W_b=(m_m - V_m\rho_w) g$ with $m_m = 77$\,mg is the mass of the magnet, $g$ is gravitational acceleration, $V_m$ is the volume of the magnet. We have $V_m \rho_w = 14$\,mg. Thus, $W_b = 
6.2 \times 10^{-4}$\,N. From the observation of $U_r$ versus $B$, we find that the robot moves forward when $A_r > 2.5$\,mm, and thus infer that $\mu_k \approx 0.005$. 
When we estimate sliding friction by tilting the substrate, we find $\mu_k \approx 0.35 \pm 0.05$ which is two orders of magnitude larger. Thus, it appears that the robot's head rolls from side to side as it advances forward. Plotting $D_s$, the substrate drag in Fig.~\ref{fig:Licomp}(a), we note that it is much smaller than the other drag components except for $f < 2$\,Hz, and can account for the finite frequency at which forward motion occurs. 

\begin{figure}
	\includegraphics[width=8.5cm]{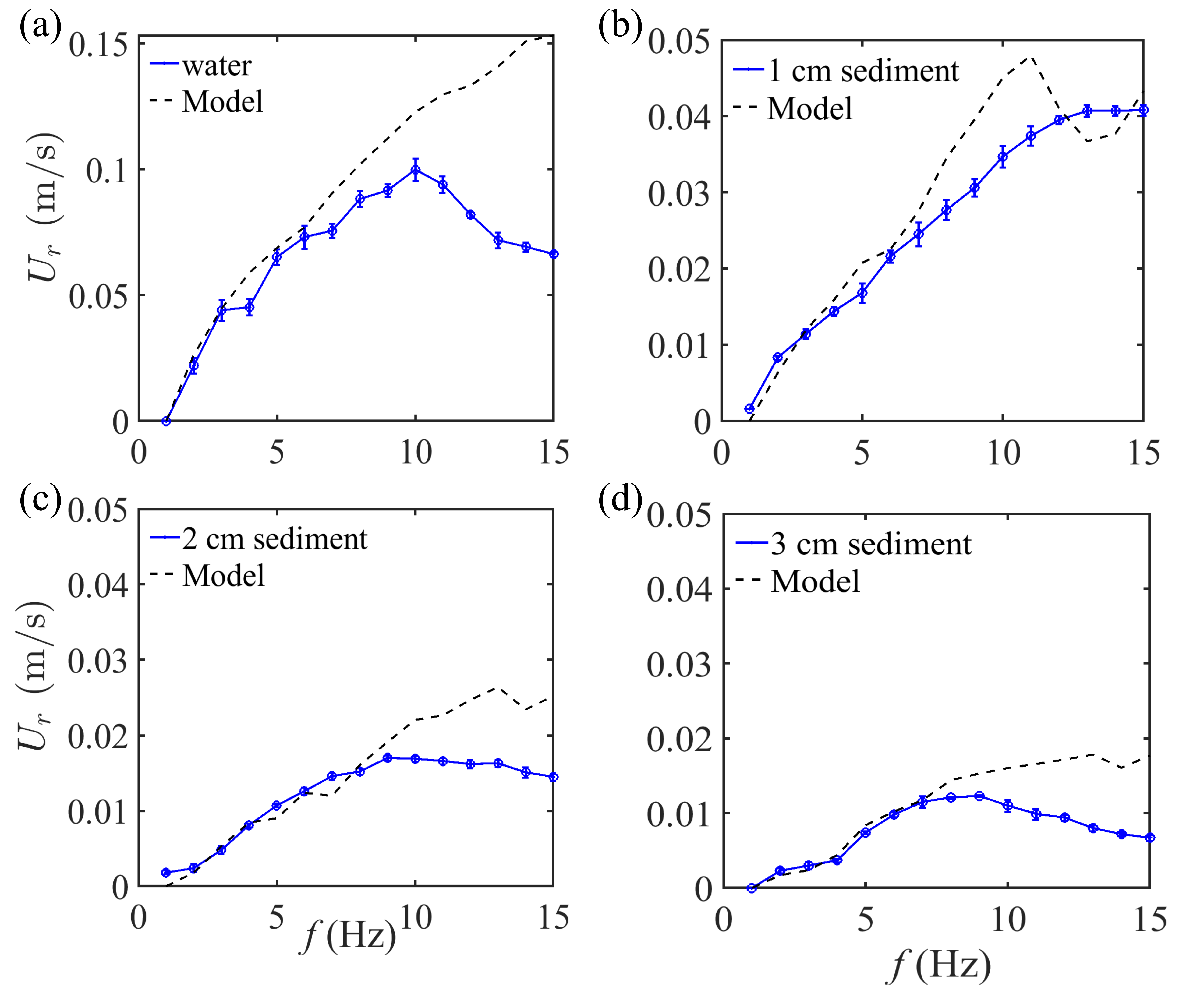}
	\caption{Comparison of the measured speeds with the model in water (a), and sediment beds (b-d). Good agreement is observed at lower $f$ with systematic deviations above $f=10$\,Hz. The volume fraction $\phi$ is used as a fitting parameter to evaluate the effective viscosity using Eq.~\ref{eq:kd}, and found to be $\phi = 0.53$ (b), $0.54$ (c), and $0.56$ (d) in the case of $h_b = 1$\,cm, 2\,cm, and 3\,cm, respectively.}
	\label{fig:Lspeed}
\end{figure}

To gauge how well Eq.~(\ref{eq:fbalance}) works, we use it to calculate $U_r$, rather than evaluating each component separately. We plot these estimated $U_r$ along with the measured speeds in Fig.~\ref{fig:Lspeed}(a). We observe that it captures observed trends reasonably well including the finite frequency at which swimming occurs, until about $f \approx 10$\,Hz. It appears that the increasing thrust is overestimated with increasing $f$ leading to the discrepancy. 

In the case of the sediments, $T_r$ and $D_s$ can be evaluated, and are plotted in Fig.~\ref{fig:Licomp}(b-d) for the three sediment heights. However, we cannot evaluate $D_h$ and $D_b$ without knowledge of $Re$ which is needed to estimate $C_D$. Now, $Re$ depends on the effective viscosity of the medium experienced by the robot which depends on the volume fraction $\phi$ according to the empirical Krieger-Dougherty relation~\cite{krieger59},
\begin{equation}
    \eta_m = \eta_w (1 - \frac{\phi}{\phi_c})^{-2.5 \phi_c}
    \label{eq:kd}
\end{equation}
where $\phi_c = 0.63$, the volume fraction at which viscosity of the sediment medium diverges. While we can obtain $\phi$ in regions behind the robot, it is difficult to estimate $\phi$ right near the robot body. Thus, we use it as a fitting parameter, and use it as a single parameter fit to obtain $U_r$ using Eq.~(\ref{eq:fbalance}) as a function of $f$ in the comparison with measured data in Fig.~\ref{fig:Lspeed}(b-d). 

For self-consistency, we plot the implied $Re$ in Fig.~\ref{fig:Re}(b-d) for the three sediment beds. We note that $Re$ are much lower than in the case of water. Nonetheless, we note that $Re \gg 1$ in case of the body, and thus adopting Lighthill's model for evaluating thrust using Eq.~(\ref{eq:thrust_alt}) may be still reasonable. We also plot the resulting head and body drag $D_h$ and $D_b$ in Fig.~\ref{fig:Licomp}(b-d) using these lower $Re$ which result in higher $C_d$. We note that head drag is systematically higher compared to body drag. And at least till $f \approx 10$\,Hz, it increases in accord with increasing $T_r$, but then falls, giving rise to the overall overestimate of $U_r$ in the sediments when $f$ increases above 10\,Hz. These deviations are clearly not captured by our simplified modeling of the forces due to the medium on the robot.

\section{Conclusions}
We constructed magnetoelastic robots with elastic bodies and demonstrated that they can move in water-saturated sediment beds with transverse undulation of the body actuated by an oscillating magnetic field. In our study, the applied oscillating field frequency, magnetic field strength, and sediment depth are control variables. A transition from nearly rigid to anguilliform body motion is observed because of subtle balance of thrust, drag, and elastic forces. Following intuition, the stroke amplitudes increase with applied field, and decrease with applied frequency due to the medium drag. By contrast, the robot speed shows a more nuanced behavior. While the speed decreases with increasing sediment depth as may be expected due to increasing drag, the speed increases initially with frequency, and then decreases after reaching a peak. The peaks depend on the sediment depth. Such a trend can arise due to the different contributions of increasing frequency and decreasing amplitude on locomotion speeds. We measured the Reynolds number by utilizing the speed of the robot, body, and tail amplitude measurements. This analysis demonstrates that the robot performs swimming in the inertial hydrodynamic regime. We also computed the thrust and drag coefficients and forces to assess their impact at the specified oscillation frequency for varying sediment heights. Given the body speeds and the highly fluidized nature of the medium, we focus on locomotion strategies which are based on inertial hydrodynamics.  

We appear to find that the overall trends in the robot speed at least in terms of the order of magnitude are generally consistent with Lighthill's theory of elongated body swimming, over a range of driving frequencies. This theory can be used to understand the important force components while swimming and burrowing in sediment medium over a range of applied field strengths and frequencies. However, there are systematic deviations in the measured speeds in water as well as in sediments are not captured by the model with increasing driving frequency. It is possible that these deviations arise due to the flow geometry in the medium in our system, whereby the flow is confined close to the robot. Thus, further elemental modeling efforts are needed to fully capture the robot forms and speeds observed in our study.

\begin{acknowledgments}
We thank Julien Chopin, Brian Chang, and Ankush Pal for stimulating discussions. This work was supported by National Science Foundation under Grant No. CBET-1805398.
\end{acknowledgments}   


\begin{thebibliography}{10}

\bibitem{barnes2009fundamentals}
Richard Stephen~Kent Barnes and Kenneth~Henry Mann.
\newblock {\em Fundamentals of aquatic ecology}.
\newblock John Wiley \& Sons, 2009.

\bibitem{lauga09}
Eric Lauga and Thomas~R Powers.
\newblock The hydrodynamics of swimming microorganisms.
\newblock {\em Rep. Prog. Phys}, 72:096601, 2009.

\bibitem{maladen09}
Ryan~D Maladen, Yang Ding, Chen Li, and Daniel~I Goldman.
\newblock Undulatory swimming in sand: subsurface locomotion of the sandfish
  lizard.
\newblock {\em Science}, 325:314--318, 2009.

\bibitem{hosoi15}
A.~E. Hosoi and D.~I. Goldman.
\newblock Beneath our feet: Strategies for locomotion in granular media.
\newblock {\em Annual Review of Fluid Mechanics}, 47:431--453, 2015.

\bibitem{maladen2011mechanical}
Ryan~D Maladen, Yang Ding, Paul~B Umbanhowar, Adam Kamor, and Daniel~I Goldman.
\newblock Mechanical models of sandfish locomotion reveal principles of high
  performance subsurface sand-swimming.
\newblock {\em Journal of The Royal Society Interface}, 8(62):1332--1345, 2011.

\bibitem{shimada2009swimming}
Takashi Shimada, Dirk Kadau, Troy Shinbrot, and Hans~J Herrmann.
\newblock Swimming in granular media.
\newblock {\em Physical Review E}, 80(2):020301, 2009.

\bibitem{ruhs2021complex}
Patrick~A R{\"u}hs, Jotam Bergfreund, Pascal Bertsch, Stefan~J Gst{\"o}hl, and
  Peter Fischer.
\newblock Complex fluids in animal survival strategies.
\newblock {\em Soft Matter}, 17(11):3022--3036, 2021.

\bibitem{gray53}
J.~Gray.
\newblock Undulatory propulsion.
\newblock {\em Journal of Cell Science}, s3-94:551--578, 1953.

\bibitem{jung10}
Sunghwan Jung.
\newblock {\it Caenorhabditis elegans} swimming in a saturated particulate
  system.
\newblock {\em Physics of Fluids}, 22:031903, 2010.

\bibitem{dorgan13}
Kelly~M. Dorgan, Chris~J. Law, and Greg~W. Rouse.
\newblock Meandering worms: mechanics of undulatory burrowing in muds.
\newblock {\em Proceedings of the Royal Society B}, 280:20122948, 2013.

\bibitem{dorgan15}
Kelly~M. Dorgan.
\newblock The biomechanics of burrowing and boring.
\newblock {\em Journal of Experimental Biology}, 218:176--183, 2015.

\bibitem{Kudrolli2019}
Arshad Kudrolli and Bernny Ramirez.
\newblock Burrowing dynamics of aquatic worms in soft sediments.
\newblock {\em Proceedings of the National Academy of Sciences},
  116(51):25569--25574, 2019.

\bibitem{herrel2011burrowing}
Anthony Herrel, Hon~Fai Choi, Elizabeth Dumont, Natalie De~Schepper, Bieke
  Vanhooydonck, Peter Aerts, and Dominique Adriaens.
\newblock Burrowing and subsurface locomotion in anguilliform fish: behavioral
  specializations and mechanical constraints.
\newblock {\em Journal of Experimental Biology}, 214(8):1379--1385, 2011.

\bibitem{yaqoob2023mechanics}
Basit Yaqoob, Andrea Rodella, Emanuela Del~Dottore, Alessio Mondini, Barbara
  Mazzolai, and Nicola~M Pugno.
\newblock Mechanics and optimization of undulatory locomotion in different
  environments, tuning geometry, stiffness, damping and frictional anisotropy.
\newblock {\em Journal of the Royal Society Interface}, 20(199):20220875, 2023.

\bibitem{Fischer2022}
Peter {Fischer}.
\newblock {Sand and mucus: A toolbox for animal survival}.
\newblock {\em Physics Today}, 75(12):30--37, December 2022.

\bibitem{suarez2003hyperactivated}
SS~Suarez and H-C Ho.
\newblock Hyperactivated motility in sperm.
\newblock {\em Reproduction in domestic animals}, 38(2):119--124, 2003.

\bibitem{spagnolie2023swimming}
Saverio~E Spagnolie and Patrick~T Underhill.
\newblock Swimming in complex fluids.
\newblock {\em Annual Review of Condensed Matter Physics}, 14:381--415, 2023.

\bibitem{Ramananarivo2013}
Sophie Ramananarivo, Ramiro Godoy-Diana, and Benjamin Thiria.
\newblock Passive elastic mechanism to mimic fish-muscle action in anguilliform
  swimming.
\newblock {\em Journal of The Royal Society Interface}, 10(88):20130667, 2013.

\bibitem{wang2021effect}
Tianlu Wang, Ziyu Ren, Wenqi Hu, Mingtong Li, and Metin Sitti.
\newblock Effect of body stiffness distribution on larval fish--like efficient
  undulatory swimming.
\newblock {\em Science Advances}, 7(19):eabf7364, 2021.

\bibitem{Naclerio2021}
Nicholas~D. Naclerio, Andras Karsai, Mason Murray-Cooper, Yasemin Ozkan-Aydin,
  Enes Aydin, Daniel~I. Goldman, and Elliot~Wright Hawkes.
\newblock Controlling subterranean forces enables a fast, steerable, burrowing
  soft robot.
\newblock {\em Science Robotics}, 6, 2021.

\bibitem{Das2023}
Riddhi Das, Saravana~Prashanth Murali~Babu, Francesco Visentin, Stefano Palagi,
  and Barbara Mazzolai.
\newblock An earthworm-like modular soft robot for locomotion in multi-terrain
  environments.
\newblock {\em Scientific Reports}, 13:1571, 01 2023.

\bibitem{naclerio2018soft}
Nicholas~D Naclerio, Christian~M Hubicki, Yasemin~Ozkan Aydin, Daniel~I
  Goldman, and Elliot~W Hawkes.
\newblock Soft robotic burrowing device with tip-extension and granular
  fluidization.
\newblock In {\em 2018 IEEE/RSJ International Conference on Intelligent Robots
  and Systems (IROS)}, pages 5918--5923. IEEE, 2018.

\bibitem{mehta2007granular}
Anita Mehta and Sam Edwards.
\newblock {\em Granular physics}.
\newblock Cambridge University Press, 2007.

\bibitem{balmforth14}
Neil~J. Balmforth, Ian~A. Frigaard, and Guillaume Ovarlez.
\newblock Yielding to stress: Recent developments in viscoplastic fluid
  mechanics.
\newblock {\em Annual Review of Fluid Mechanics}, 46:121--146, 2014.

\bibitem{panaitescu17}
A.~Panaitescu, X.~Clotet, and A.~Kudrolli.
\newblock Drag law for an intruder in granular sediments.
\newblock {\em Physical Review E}, 95:032901, 2017.

\bibitem{nichol2010flow}
Kiri Nichol, Alexey Zanin, Renaud Bastien, Elie Wandersman, and Martin van
  Hecke.
\newblock Flow-induced agitations create a granular fluid.
\newblock {\em Physical review letters}, 104(7):078302, 2010.

\bibitem{jewel18}
R.~Jewel, A.~Panaitescu, and A.~Kudrolli.
\newblock Micromechanics of intruder motion in wet granular medium.
\newblock {\em Physical Review Fluids}, 3:084303, 2018.

\bibitem{allen19}
Benjamin Allen and Arshad Kudrolli.
\newblock Effective drag of a rod in fluid-saturated granular beds.
\newblock {\em Phys. Rev. E}, 100:022901, 2019.

\bibitem{Chang2022}
Brian Chang and Arshad Kudrolli.
\newblock Nonadditive drag of tandem rods drafting in granular sediments.
\newblock {\em Phys. Rev. E}, 105:034901, Mar 2022.

\bibitem{lighthill1970}
M.~J. Lighthill.
\newblock Aquatic animal propulsion of high hydromechanical efficiency.
\newblock {\em Journal of Fluid Mechanics}, 44(2):265–301, 1970.

\bibitem{pal2021}
Ankush Pal and Arshad Kudrolli.
\newblock Drag anisotropy of cylindrical solids in fluid-saturated granular
  beds.
\newblock {\em Phys. Rev. Fluids}, 6:124302, Dec 2021.

\bibitem{Supplemental}
See supplemental material at. the videos show the top and side view of the
  robot's motion in sediment media.

\bibitem{lighthill76}
James Lighthill.
\newblock Flagellar hydrodynamics.
\newblock {\em SIAM Review}, 18:161--230, 1976.

\bibitem{Wang2016}
X.~Wang and C.~Hopkins.
\newblock Bending, longitudinal and torsional wave transmission on
  euler-bernoulli and timoshenko beams with high propagation losses.
\newblock {\em The Journal of the Acoustical Society of America},
  140(4):2312--2332, 2016.

\bibitem{tritton59}
D.~J. Tritton.
\newblock Experiments on the flow past a circular cylinder at low reynolds
  numbers.
\newblock {\em Journal of Fluid Mechanics}, 6(4):547--567, 1959.

\bibitem{krieger59}
I.~M. Krieger and T.~J. Dougherty.
\newblock A mechanism for non-newtonian flow in suspensions of rigid spheres.
\newblock {\em Transactions of the Society of Rheology}, 3:137, 1959.

\end{thebibliography}

\end{document}